\documentclass[12pt]{iopart}

\usepackage{graphicx}
\usepackage{amssymb}
\usepackage{epstopdf}
\usepackage{subfig}
\usepackage{fancyhdr}

\usepackage{enumitem}

\mathchardef\mhyphen="2D
\begin{document}

\title{The case for absolute ligand discrimination : modeling information processing and decision by immune T cells}
\author{Paul Fran\c cois$^1$, Gr\'egoire Altan-Bonnet $^2$}
\address{$^1$ Ernest Rutherford Physics Building, McGill University, H3A2T8 Montreal QC, CANADA}
\address{$^2$ Program in Computational Biology \& Immunology, Memorial Sloan Kettering, New York NY, USA}

\maketitle

\section{Introduction}

Cells within the body are constantly bombarded with a large repertoire of molecules that must be dealt with as potential stimuli. Most of the time, these molecular inputs are measured by receptors at the surface of the cells. State of these receptors are thus informative on the outside world, and experimental and theoretical biophysicists~\cite{Tkacik:2011} have extensively used information theory to estimate how much information (in the Shannon sense~\cite{SHANNON:1948}) can be encoded  (see {\it e.g.} in a developmental context work of Gregor and coworkers~\cite{Gregor:2007}).

Over longer time-scales, information processing eventually leads to a decision, that we define as a change in physiological macroscopic behavior or in the steady states of a gene regulatory network. Decision is  by essence a computation process, based on sometimes limited information. For instance, in the well-studied example of bacterial chemotaxis, a cell might decide to switch behavior between tumbling or swimming~\cite{Block:1983}, a strategy that can be well-explained  by a Max/Min game theory model~\cite{Celani:2010}.  
It is not even always desirable to optimize information collection if the environment changes too rapidly, as illustrated by the``info taxis" strategy~\cite{Vergassola:2007}. Other examples include cellular commitment to a given fate in response to dynamical signaling pathways~\cite{Corson:2012}, or decision to take action in the case of immune responses, characterized by binary Erk phosphorylation, cytokine release, and cell proliferation~\cite{AltanBonnet:2005}. The later decisions are irreversible, indicating that computation is accompanied by  information erasure and thus energy dissipation \cite{Mehta:2012ji}. Additionally, information processing may be multi-tiered in order to retrieve different features of the input stimuli: rapid decision could discriminate between ligands of different nature, while slower decision could report the quantity of ligands.

In this review, we will focus on one specific cellular decision where similar considerations apply: the ability of T cells to discriminate very specifically between self and not-self ligands. The functional significance of such ligand discrimination is quite obvious. If a T cell ``recognizes" a (potentially single) ligand as foreign, a large set of responses is triggered to eradicate the pathogenic infection that generated this stimuli. Conversely, if a T cell interacts with (many present) self ligands as self it should remain quiescent to avoid auto-immune catastrophe.

This discrimination task is particularly daunting as T cells are constantly exposed to a large number of molecular stimuli at once. This issue of signalling pleiotropy is potentially a very generic problem in biology  and we will coin the term ``absolute discrimination" to describe it. Multitudes of receptors are indeed shared in examples as different as BMP signalling, olfaction, endocrine signalling, etc...  More theoretical works have suggested that organization of immune repertoire requires strong overlapping signals \cite{Mora:2010b,Mayer:2015ce}.  In the context of this review, absolute discrimination thus is the specific and sensitive recognition of foreign ligands, independent of ligand quantity. Absolute discrimination of massive amounts of self vs low amount of foreign ligands is expected to be challenging since some self-ligands might be very close biochemically to foreign ligands.

The first part of this review will be devoted to a formal introduction to the problem of immune recognition by T cells, presenting current experimental understanding, past and present attempts to model this decision problem, and introducing the paradigm of adaptive sorting. In the second part we will introduce our current model for absolute immune discrimination, at the cellular scale. Finally, we will discuss how tools borrowed from statistical physics are needed to understand the higher level of processing in the immune system, at the cellular population scale.

\section{Theoretical approaches for absolute ligand discrimination.}

T cells  probe their environment in search of potential foreign peptides. This is done via the interaction of their T cell receptors (TCR) with ligands (pMHC), presented by Antigen Presenting Cells. At a given time, these cells ``present" a repertoire of oligopeptides (embedded within an MHC) that is representative of the current proteome (i.e. a mix of peptides from the self genome as well as a potential genome of the pathogen). The core function of T cells is to scan such repertoire and detect the presence of pathogen-derived ligands and respond, while not responding to self-derived ligands. 

This decision must be, by essence, absolute in the sense that it must be determined by ligand quality (here self or not-self), independently of ligand quantity (i.e. how many ligands  there are). In that context,  decision has been shown to be logically all-or-none, via binary/bistable response in Erk phosphorylation~\cite{AltanBonnet:2005,Lipniacki:2008,Artomov:2007} or in NFAT translocation to the nucleus.  A first difficulty is that there is no qualitative biochemical difference (e.g. a clear-cut structural distinction) between self-derived pMHC and pathogen-derived pMHC. Another natural hypothesis would be that foreign ligands lead to specific allosteric modifications (conformational  changes) at the level of T cells receptors, which would be an ideal way to confer extreme sensitivity and specificity to immune recognition. Molecular immunology has made a lot of progress in listing all components implicated in this early response, but could not find evidence for such a direct qualitative sensing in the general case. Hence T cells must make a discrimination decision based on continuous quantitative biophysical differences between self and not-self, explaining why mathematical  and physical modeling must be called upon to address how continuous variation in ligand characteristics gets processed with absolute discrimination.

\subsection{Insight from biophysics: the lifetime dogma of antigen discrimination. }
 
 \subsubsection*{Antigen discrimination is set by the lifetime of the antigen/receptor complex.}

\begin{figure}
\begin{center}
\resizebox{\textwidth}{!}{\includegraphics{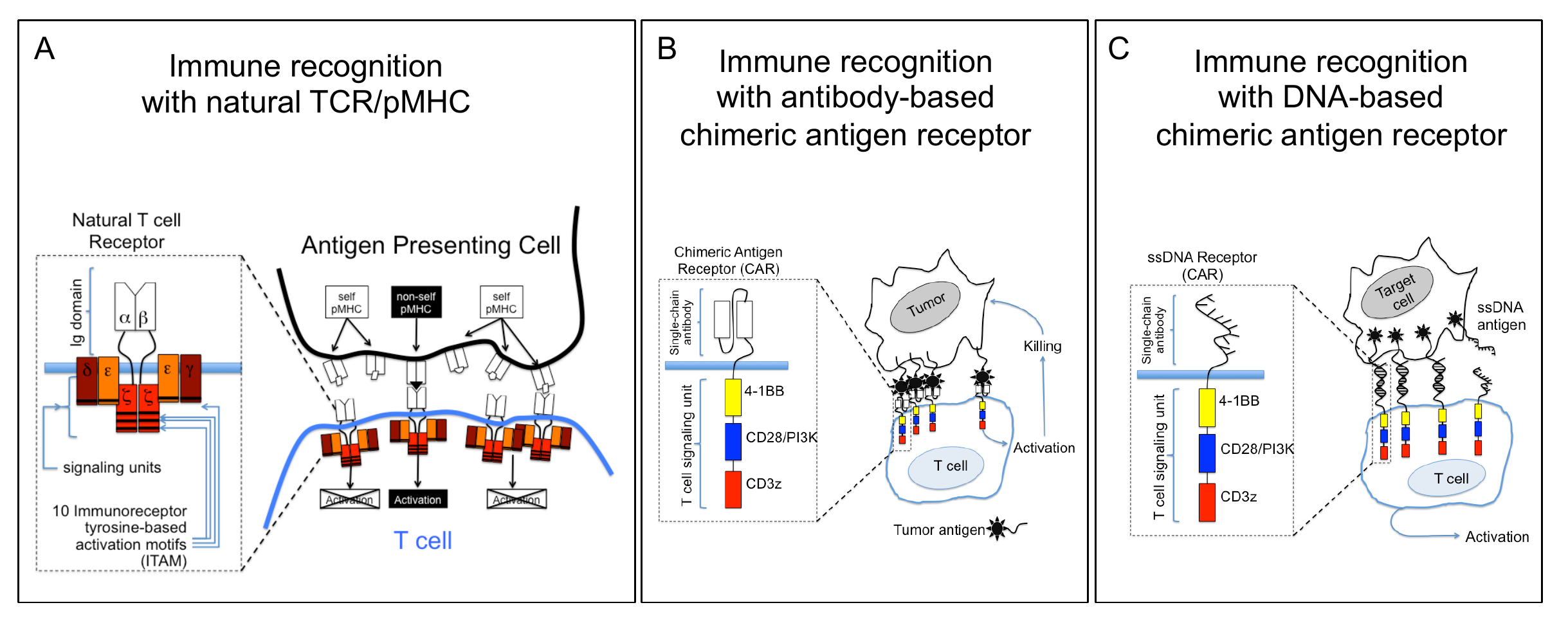}}
\end{center}
\caption{Three examples of immune recognition by T cells. {\bf A.} The natural system of antigen discrimination by T lymphocytes relies on the T cell Receptor. It is composed of an extracellular a/b domains that interact with peptide-MHC complex on the surface of antigen-presenting cells, and 6 intracellular domains containing 10 Immunoreceptor Tyrosine-based Activation Motifs (ITAMs) that get phosphorylated upon engagement with not-self ligands, and trigger T cell activation. {\bf B.} Recent developments in the field of immunotherapy introduced Chimeric Antigen Receptor (CAR): its extracellular domain is composed of a monomeric antibody that is specific for an antigen on the surface of the targeted tumor; its intracellular domain concatenates one ITAM-containing domain (z) and additional costimulatory domains (CD28 and 4-1BB) to induce robust T cell activation upon CAR engagement with its ligand. Such CAR combines the specificity of an antibody-based recognition, with robust signaling response. {\bf C.}  A DNA-based Chimeric Antigen Receptor was recently proposed by Ron Vale and coworkers~\cite{Taylor:2015}: its intracellular domain is based on the concatenation of signaling domains -similarly to the CAR described in B. Its recognition platform is composed of a single-stranded oligonucleotide, that recognizes another single-stranded oligonucleotide by sequence complementarity. Rather than relying on natural pMHC ligands whose biophysical characteristics are not tunable, Vale et al.'s design can be engineered to achieve variable lifetime for the ligand-receptor complex. Such ingenious experimental design will be critical to probe the sufficiency and limits of the lifetime dogma.}
\label{figure:Immunoreceptor} 
\end{figure}

The exact molecular events associated with self/not-self ligand discrimination by T cells remain elusive. However, immunologists, structural biologists and biophysicists have made great progress to extract key parameters that physicists can build upon to tackle the issue of specific immune sensing (Figure~\ref{figure:Immunoreceptor}.A). 

The first insight came in the 90s when researchers measured the biophysical characteristics of ligand-receptor interactions using purified proteins assayed {\it in vitro} for binding/debinding ({\it e.g.} detection by surface-plasmon resonance or by calorimetry). Kersh \etal established a hierachy of ligands with similar binding activities, where the life-time of the ligand-receptor complex determines their ability to trigger response (Table 2 in \cite{Kersh:1998wx}). Qualitatively, there exists a threshold of binding time (around 3-5s) so that for ligands with a lower binding time, T cells do not respond, while for ligands with higher binding time T cells do respond (see in particular Table 2 and Figure 3 in ~\cite{Gascoigne:2001}), thus realizing absolute discrimination based on binding. Such experimentally-derived rule (so-called ``lifetime dogma"~\cite{Feinerman:2008b}) was well established by the turn of the millennium such that it became the springboard for many modeling efforts.

It should be immediately pointed out that like any dogma, this one is not absolute, and there are exceptions to the rule (see~\cite{Feinerman:2008b} for a discussion of some exceptions,  \cite{Dushek:2011} for an experimental approach on cell populations). For instance, it has been seen that some ligands could somehow ``compensate'' a small lifetime with a very high $k_{on}$ \cite{Govern:2010kx}, which has been interpreted as an effect due to constant rebinding \cite{Chakraborty:2014hw} . More recent measurements have been carried in the context of more complete biological systems (e.g. T cells reading their ligands on the surface of antigen presenting cells), with single ligand resolution: these brought about a correction on parameters for association and dissociation rates, but concurred qualitatively with the previously-acquired {\it in vitro} measurements.  

Recent work by Cheng Zhu and coworkers  has challenged the lifetime dogma,  using a Molecular Force Spectroscopy to interrogate individual pMHC--TCR interactions. Such technique relies on micropipette manipulation of pMHC-coated beads  and exquisite measurement of the force induced by the engagement with one individual TCR on the surface of T cells to resolve the dynamics of ligand--receptor engagement. Zhu {\it et al.}'s measurements lead to paradoxical results at first: strong ligands that trigger T cells were found to be weaker binder to the receptor, thus inverting the life-time dogma~\cite{Huang:2010df}.
Subsequent studies tested how force loading on the ligand-receptor complex would alter its lifetime, and the more intuitive hierarchy of ligands was recovered, with better binders inducing better signaling responses~\cite{Liu:2014}. Zhu and colleagues thus proposed a dynamics structural model, whereby agonist ligands induce a conformational change in the complex (so-called catch bonds) that triggers T cell activation. Alternatively, non-activating ligands ({\it e.g.} self-derived peptide MHC) would not induce such conformational change, would be released rapidly (so-called slip bond) and would fail to activate a significant signaling response. Hence, there would in the end be qualitative differences between activating and non-activating ligands. This result, while potentially establishing absolute ligand discrimination {\it at the structural level}, must be reconciled with the observation that non-activating ligands may be sufficient to activate T cells when proper external cues are provided in the form of cytokines (cf section~\ref{section:cooptation}). Ultimately, the mechanical differences between catch bonds and slip bonds for pMHC-TCR pairs, as uncovered by Zhu {\it et al.}, remain to be interpreted in their capacity to trigger signal transduction.

From the Physics point of view, one intriguing aspect of these force measurements would be to add mechanical aspects to ligand discrimination. Understanding the forces associated with ligand-receptor interactions and the coupling with the mechanics of membrane deformation would be critical to account for the differential potency of ligands to activate T cells. Such quantitative models have been introduced~\cite{Qi:2001}, based on Ginzburg-Landau equations coupling the biochemistry of ligand-receptor interactions with the energetic cost of membrane deformation. Such models established that biochemical/mechanical coupling could be sufficient to physically sort membrane proteins on the T:APC cell interface, and generate a threshold of activation. Such physical models generated intriguing predictions that were subsequently validated experimentally: of note, it predicted that a family of ligands (with intermediate binding capacity) would abrogate the formation of so-called immunological synapse (a self-assembled bull-eye structure at the surface of T cells, where TCR aggregates at the center the synapse, and adhesion molecules occupy the periphery of synapse). 
In our context of immune recognition, one must point out that such synapse formation occurs downstream passed the initial signaling response associated with the ligand discrimination: it may constitute a reinforcing mechanism to anchor ligand discrimination over longer timescales, rather than the core cell-decision we are focusing on in this review. Recent models have explored how membrane stiffness influences effective binding times via supradiffusive effects ~\cite{Allard:2012}.

Two additional lines of work must be added to the biophysical conundrum of self/not-self ligand discrimination by T cells. First, in the field of immunotherapy, researchers have engineered T cells with synthetic chimeric-antigen receptors (CAR) whose extracellular domain is composed of an antibody recognizing a protein on the surface of tumors to be targeted (e.g. CD19 for B cell lymphoma), and whose intracellular domain is derived from signaling components of T cells (Figure~\ref{figure:Immunoreceptor}.B): engagement of these receptors (with non-physiological ligands of surface antigens with very large lifetime) has been shown to be necessary and sufficient to activate T cells. In fact, examples of supra-physiological lifetimes for antigen/receptor complexes that lead to T cell activation were derived experimentally by in vitro evolution of the TCR/pMHC complex~\cite{Holler:2000}.  In the context of modeling early immune detection, this is relevant as the biophysics of ligand-receptor interaction are very different (with very large binding affinities), yet consistent with the lifetime dogma: antigen/receptor pairs with very strongly-held complexes, and very large lifetimes are indeed very stimulatory. 

Another line of experimental evidence has recently been reinforcing the lifetime dogma. Markus Taylor \& coworkers~\cite{Taylor:2015} engineered a new class of chimeric antigen receptors, whose extracellular recognition unit is composed of single-stranded DNA \ref{figure:Immunoreceptor}.C). Antigens for these T cells are composed of complementary single-strands of DNA (e.g. an oligomer of adenosines and cytokines, to avoid secondary structures). Hence immune detection in that context is highly tunable, quantifiable and easy to model: it is essentially the biophysics of DNA hybridization that drives the engagement of this artificial antigen receptor. In that context, Taylor et al. demonstrated that the association rates of these artificial receptor/ligand pair were essentially constant as it is limited by the nucleation of double stranding between two complementary DNA pairs. However, the dissociation rates are highly variable and essentially dominated by the free energy of double-strand formation. Hence, these DNA-based chimeric antigen receptor and ligands recapitulate the biophysical characteristics of ligand-receptor interaction in the natural immune detection context. Most strinkingly, Taylor et al. found that the lifetime dogma holds with a threshold of activation set around 3s for the lifetime of the antigen-receptor complex~\cite{Taylor:2015}.

As of 2015, although the structural details of the early events in immune recognition by T cells remain elusive, the consensus around the lifetime dogma is thus holding and it is enabling physicists to build biochemically-explicit or phenotypic models of good biological significance~\cite{AltanBonnet:2005,Francois:2013,Lever:2014}. It constitutes a rich paradigm for both theoretical and experimental biophysical considerations, and most of our discussion will be within this framework. 

\subsection{Setting the problem for physicists:  what does absolute immune discrimination entails ?}

\begin{figure}
    \begin{center}
\resizebox{\textwidth}{!}{\includegraphics{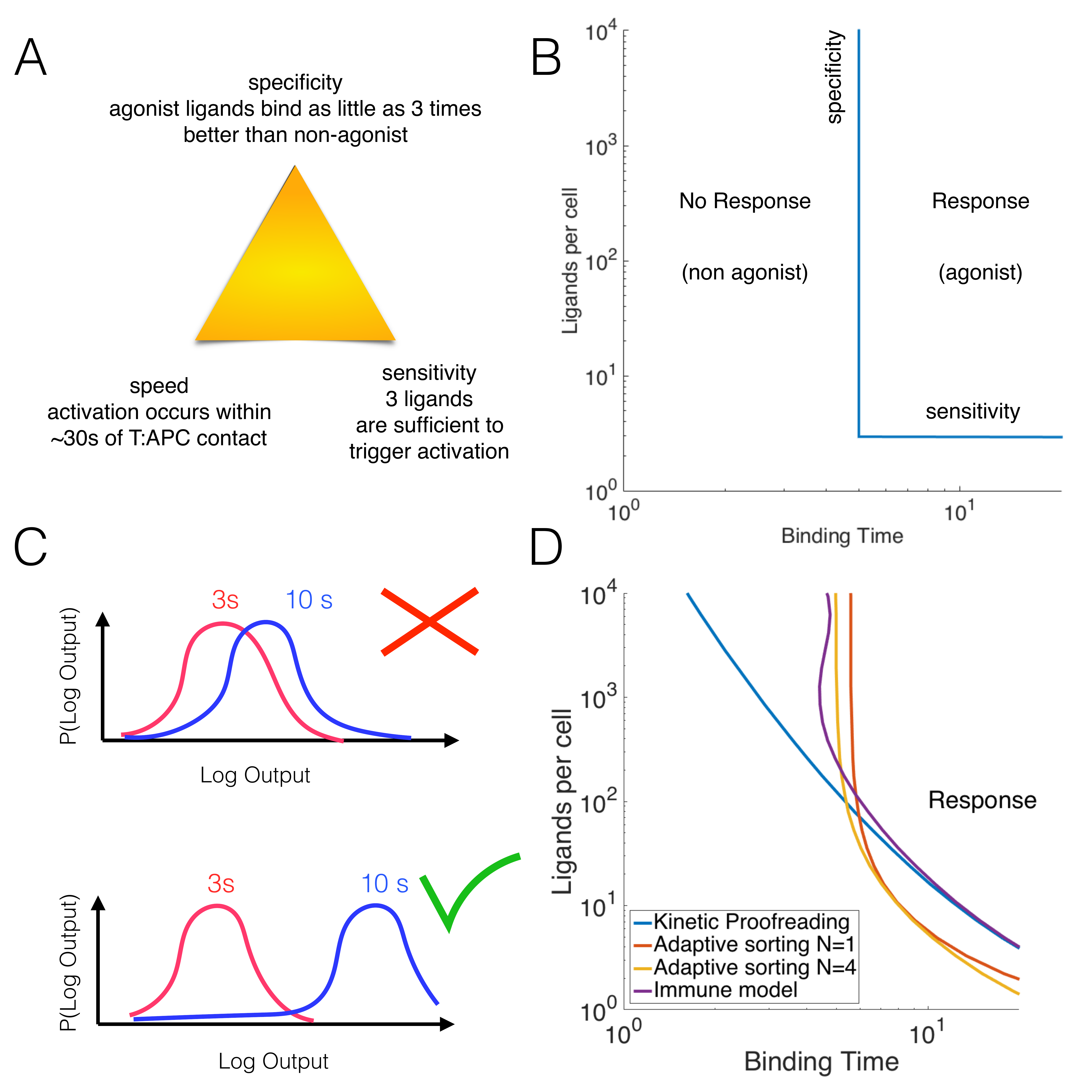}}
    \end{center}
    \caption{Absolute discrimination for physicists. A: Golden triangle for immune recognition. B: Idealized response line corresponding to the immune golden triangle. C: Two possible distributions for an Output variable directing immune decision, for networks exposed to random concentrations of ligands with identical binding times. In the top example, typical values for ligands with $\tau=3 \ s$ and $\tau=10 \ s$ overlap, so that it is not possible to discriminate between these ligands. In the bottom example, those distributions are well separated so that it is possible to choose a thresholding procedure on this Output to ensure absolute discrimination. D: Typical response lines for various models discussed in this review. Decision threshold was adjusted to have $\tau_c \sim 5 \ s$. Parameters for KPR and immune model from \cite{Francois:2013}, and from \cite{Lalanne:2013} for adaptive sorting } \label{figure:absolute} 
\end{figure}

In recent years, quantitative immunology has partially characterized the ``phenotypic space" of T cells as a function of these parameters. A ``golden triangle" characterizing immune response can be drawn~\cite{Feinerman:2008b}, Figure~\ref{figure:absolute} A. The first vertex of this triangle is {\it ligand specificity}, as encapsulated in the lifetime dogma described in the previous section: there exists an absolute discrimination threshold on ligand binding time, around 3-5 s.

The second vertex of the triangle is {\it ligand sensitivity}. Minute amounts of ligand are able to trigger response. Actually, there are strong experimental evidence that one foreign ligand can trigger immune response~\cite{Huang:2013}, so that the physical limit of detection is reached biologically, a situation reminiscent of other famous examples such as photon sensing~\cite{Bialek:2012}. Such high sensitivity might be functionally critical as the immune system can ``snip" a pathogenic infection before it has a chance to expand.

The last vertex of this triangle is {\it decision speed}. We know from experiments that immune decision at the single cell level is taken within a couple of minutes ~\cite{Zell:2001}. Note that this decision time depends quite strongly on ligand concentration~\cite{AltanBonnet:2005} yet it is relatively fast to accommodate the limited time T cells spend scanning the surface of one antigen--presenting cell. 

To reformulate this problem  in a generic way, imagine a cell with a given set of identical receptors is suddenly exposed to $L$ ligands, with identical  binding time $\tau$. We can plot in the $(L,\tau)$ plane a ``response line", characterizing the boundary between responding regions (``agonist"  ligands) and non-responding ones (``non-agonist" ligands). The life-time dogma states that below some critical time $\tau_c$, for all $L$, there is no response, while above $\tau_c$, for any $L$ there is response.    This defines a vertical line in the $(L, \tau)$ plane. In biological terms, this is usually called a ``specific'' response to a category of ligands, and thus we call this line ``specificity line''. This line corresponds to the first vertex of the golden triangle.

  It should be noted immediately that it is not obvious how such response can be realized, especially it seems a priori impossible to have such specific response for very small $L$ (we can not have response without signal !). Indeed, in any kind of biological settings, an obvious physical limitation is that there can not be less than $1$ ligand presented at any given time. Immune cells can nevertheless trigger response when exposed to $1$ to $3$ agonist ligands: this defines a (horizontal) sensitivity line. The corresponding idealized  lifetime dogma response line is displayed on Figure~\ref{figure:absolute} B. A third dimension would be necessary to account for the third element of this triangle, speed, but is not drawn here.

\subsection{Early attempts at modeling immune recognition: kinetic proofreading}

Historically, McKeithan~\cite{Mckeithan:1995} was the first to propose a mechanistic model to underly the early events in immune recognition, accounting for such qualitative ligand recognition. The control of the quality of immune response by a single kinetic  parameter $\tau$ is reminiscent of the famous Hopfield-Ninio kinetic-proofreading (KPR) paradigm, first proposed in the context of DNA replication and protein translation~\cite{Hopfield:1974,Ninio:1975}. In the immune context, McKeithan pointed out that subsequent to TCR-pMHC interactions, the receptor does internally go through several rounds of phosphorylation.  Calling $C_n$ the ligand-receptor complex that has reached the $n^{th}$ degree of phosphorylation in the cascade, $L$ the ligand and $R$ the receptor, simplified equations for a continuous model of this process are:
\begin{eqnarray}
\dot C_0=&-(\phi+\tau^{-1} ) C_0+\kappa (L-\sum_{i} C_i) (R-\sum_{i} C_i) \label{KPR:eq-1} \\
\dot C_n=&-(\phi+\tau^{-1} ) C_n+\phi C_{n-1}, \hspace{0.15cm} 1<n<N\\
\dot C_N=&- \tau^{-1}  C_N+\phi C_{N-1} \label{KPR:eq-3} 
\end{eqnarray}
where $\kappa$ is the association rate between ligand and receptor and $\phi$ the phosphorylation rate in the cascade (here and in the following, we make a mean field approximation and concentrations are measued in units of ``molecule per cell''). Parameter $\tau$ corresponds to binding time of ligand to the receptor. It appears in equations for all $C_n$s, accounting for the hypothesis that after unbinding of the pMHC ligand, receptor would be quickly dephosphorylated and the phosphorylation cascade would need to restart from ``ground zero" (i.e. $C_0$).

At steady state, assuming $R$ is far from saturation and that $\phi<<\tau^{-1}$, one can easily derive that the last complex $C_N$ thus has concentration scaling as $L \tau^{N+1}$: this is the usual geometric dependency characteristic of kinetic proof--reading with $N$ steps. Its role is to amplify difference between ligands:  with only 2 steps,  for the same initial ligand concentration, one can get up to 6 orders of magnitude in the difference of concentration of $C_N$ for self ligands ($\tau=0.1 \quad s$ ) vs agonist ligand ($\tau=10  \quad s $). So if immune decision is taken via a downstream thresholding mechanism, physiological concentration of self ligands can not trigger immune responses even though one single agonist ligands theoretically can (see response line on Figure~\ref{figure:absolute}~D).

However, there are several quantitative shortcomings for a simple proofreading model. First, it is well known that to work efficiently, KPR needs to be very slow, which renders it incompatible with the observed  fast response times of adaptive immune responses~\cite{AltanBonnet:2005}. Second, the response line of KPR for any reasonable number of proofreading steps with realistic decision speed simply does not account for the observed specificity in terms of binding time $\tau$, as illustrated on Figure~\ref{figure:absolute}~D. Finally, mixtures of ligand with different binding times will yield purely additive response to a kinetic proofreading mechanism, and thus would not account for more puzzling (but yet fundamental) aspects of immune response such as antagonism, where some non-reactive ligands can actually inhibit response of not self ligands ~\cite{AltanBonnet:2005}.

Thus, one needs to augment the traditional KPR scheme with feedback regulation in order to be able to account for our ``golden triangle" (specificity, sensitivity and speed) that characterizes the early events of T cell activation.

\subsection{In silico evolution and adaptive sorting}

The most counter-intuitive and puzzling property of immune response is its high specificity. The reason is that we would expect a priori that some shorter binding time could be ``compensated" by higher ligand concentrations. More quantitatively, similar to the kinetic proofreading model, one would expect in general that any output $O$ of a general signaling would behave as $O=f(L,\tau)$, where $f$ is a monotonic function of both $L, \tau$. But then how could we have a sharp process so that, on the response curve, a small decrease in $\tau$ (from agonist to self) leads to a change of $L$ or several orders of magnitude ? 

To answer this question, we turned to {\it in silico evolution}~\cite{Lalanne:2013} (a review of this method can be found in~\cite{Francois:2014}) .
 The idea is to simulate a Darwinian process on a space of possible models to select for absolute discrimination. Considering a population of biochemical networks (typically 30), our algorithm randomly mutates networks. Possible evolutionary moves are inspired by proofreading-based models of immune recognition, and consist in addition/removal of proofreading steps, as well as addition/removal of internal phosphorylations and dephosphorylations, or of kinase/phosphatases. For selection, we need to define a scoring or ``fitness" function, and we chose to use mutual information as explained below.
 
  Assume a cell is exposed to one type of ligands, binding time $\tau$, with probability $p_\tau(L)$. We want discrimination to be efficient over several orders of magnitude in  $L$ concentration. In the absence of any other information we choose $p_\tau(L)$ to be uniform on a log scale, within physiological concentration range. Such a choice is also consistent with the well-known fact that concentrations in cells are distributed log-normally \cite{AltanBonnet:2005}.
  
  Consider an Output variable $O$ \footnote{the nature of $O$ is itself under selective pressure in the algorithm so that evolution can choose the variable carrying maximum information for discrimination as $O$ }.To each couple $(L,\tau)$ corresponds a distribution of output variable $O$ characteristic of the signalling pathway, that we call $p_\tau(O|L)$ . Since our problem for absolute ligand discrimination implies immune detection independently of ligand concentration, let us marginalize over all possible ligand concentrations and define a probability distribution for this Output associated to a binding time $\tau$ :
\begin{equation}
P_\tau(O)=\int p_\tau(O|L)p_\tau(L) dL
\end{equation}

This probability distribution is then a pure function of the binding time of the ligand. Good ligand discrimination will be possible only if there is very little overlap between distributions corresponding to different $\tau$s (see Figure~\ref{figure:absolute}~C for an example of distributions with two different ligand types).

A practical way to use this for {\it in silico evolution} is to consider a situation where cells can be exposed to two different types of ligands (say $\tau=3s$ and $\tau=10s$ to fix ideas).  To maximize selection for efficient discrimination, we assume equal probability for observing these two $\tau$s. Then, based on computed distributions such as the ones on  Figure~\ref{figure:absolute}~C,  we compute mutual information between the Output and $\tau$, based on  $P_\tau(O)$. A mutual information of $1$ bit means that perfect discrimination is possible. 

\begin{figure}
    \begin{center}
\resizebox{\textwidth}{!}{\includegraphics{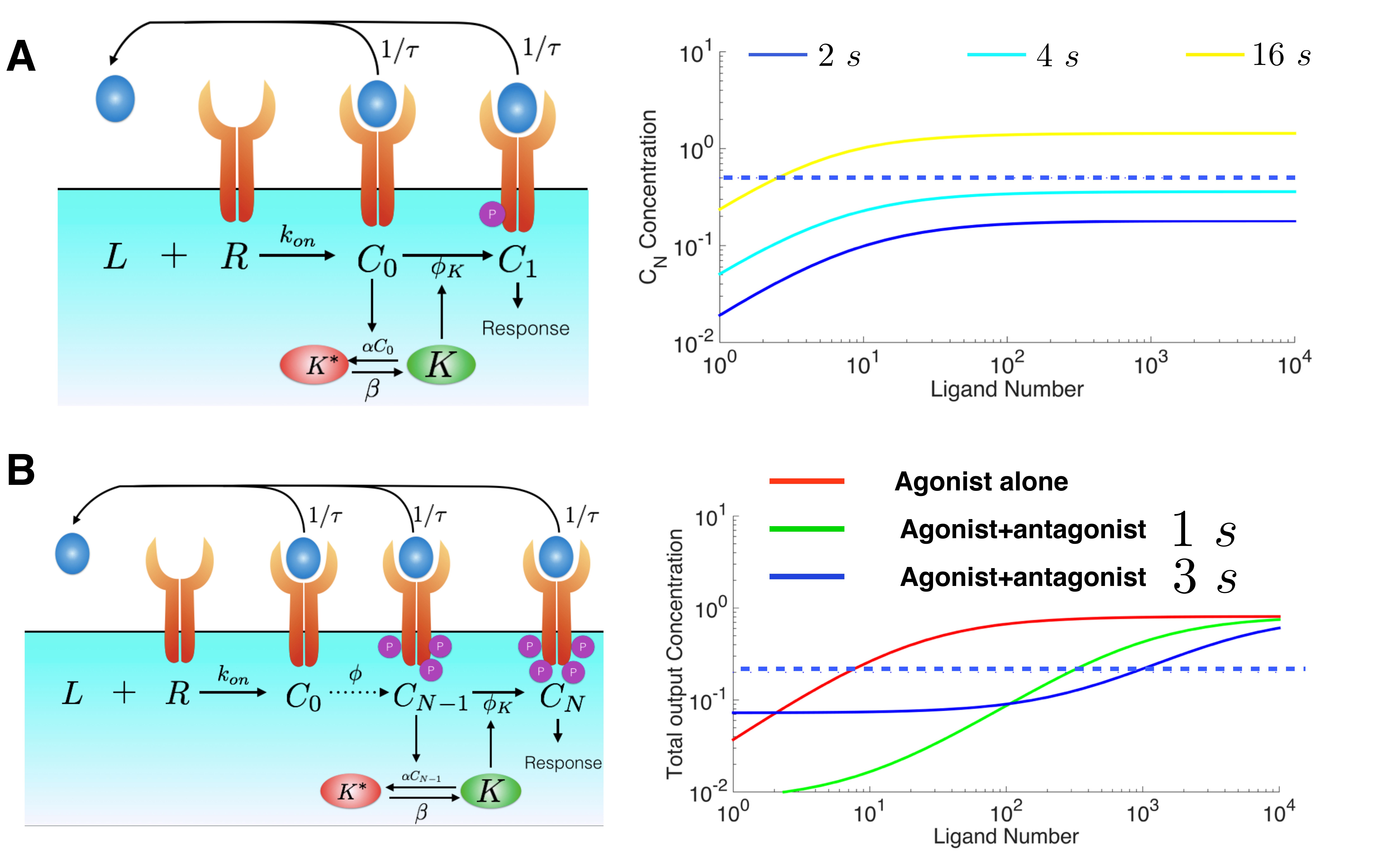}}
    \end{center}
    \caption{ {\it Adaptive sorting} networks. (A) One-step adaptive sorting topology evolved in~\cite{Lalanne:2013}. Output concentration as a function of ligand for different values of $\tau$ is represented. Dashed blue lines indicate thresholding for decision for deterministic systems used for simulations. (B) Adaptive sorting with more proofreading steps. Output concentration for agonist  ($\tau=10 \ s$)  mixed with  and $10^4$ sub thresholds ligands illustrates antagonism (response for pure ligands is qualitatively identical to the network of panel A and thus is not shown)  } \label{figure:AS} 
\end{figure}

This evolutionary procedure quickly converges to a very simple scheme described in~\cite{Lalanne:2013}, that we called {\it adaptive sorting} (Figure \ref{figure:AS} A).  Simplified continuous equations for adaptive sorting are:

\begin{eqnarray}
\dot C_0=&\kappa (L-\sum_{i} C_i) (R-\sum_{i} C_i) -(\phi_K(C_0)+\tau^{-1} ) C_0,\\
\dot C_1=&\phi_K(C_0) C_0 - \tau^{-1}  C_1, \label{C1} \\
\phi_K(C_0)=&\frac{\phi}{C_0+C^*}  \label{AS:eq}
\end{eqnarray}

The basis of this network is a one-step kinetic proofreading process, that is modulated by regulation of the phosphorylation rate $\phi_K$ from $C_0$ to $C_1$. This term is a Michaelis-Menten function that can be interpreted as a repression by the first complex in the cascade ($C_0$) of kinase $K$ responsible for its phosphorylation.   Total contribution of phosphorylation $\phi_K(C_0) C_0$ therefore contains two $C_0$ dependency: a direct linear increasing contribution $C_0$ (substrate of the phosphorylation), and an indirect decreasing $\phi_K(C_0)$ (regulation of phosphorylation by substrate), thus encoding a so-called incoherent feedforward loop \cite{Mangan:2003}. In the limit of high $C_0$ (and thus high $L$), those $C_0$ dependencies compensate  yielding an output  at steady state from equations \ref{C1}-\ref{AS:eq}:

\begin{equation}
C_1=\tau \phi_K(C_0) C_0 \simeq \phi \tau \textrm{\quad if \quad } C_0>>C^* \label{asympt}
\end{equation}
The later expression is independent from the amount of ligand presented, and then is a pure function of binding time  as illustrated on Figure~\ref{figure:AS}~A. Any thresholding process on $C_1$ can thus efficiently discriminate between binding times. Response line of network is illustrated on Figure~\ref{figure:absolute}~D  for a simple thresholding process on $C_1$ (``Adaptive sorting N=1"), in close agreement with the idealized response from  Figure~\ref{figure:absolute}~B.

It should be stressed that there are two very important biochemical assumptions related to kinase $K$ in adaptive sorting:  \textit{i.} it should diffuse rapidly inside the cell and  \textit{ii.} it should belong to a pool shared by all receptors. Since any bound receptor can deactivate $K$, total $K$ thus aggregates global information over multiple bound receptors. $K$ in turns tune local state of receptors, so that the total activity of the kinase $K$ is a decreasing function  $\phi_K(C_0)$ of total $C_0$.  Overall, $K$ effectively couples the different receptors,  which explains the non-linearity of the output as a function of the ligand concentration.

Revisiting the golden triangle, by design, adaptive sorting realizes specificity (i.e. discriminates between different  $\tau$s) in the limit of high ligand concentration.   Sensitivity (to small ligand concentration) depends on the relative values of $C_0$ and $C^*$:  asymptotic limit  of equation \ref{asympt} is realized  only when $C_0$ is higher than $C^*$.  For lower value of $C_0$, $\phi_K$ is approximately constant and one recovers a one-step kinetic proofreading regime. So while discrimination very close to  threshold can not be fully absolute for small $L$, stronger agonists still trigger response at low concentration as indeed visible on Figure~\ref{figure:absolute}~D.  Speed is related to low number of molecules: considering the immune example, if a T cell is able to perform detection for ligand concentration as low as one ligand per cell, then one would expect a potentially deleterious sensitivity to stochastic fluctuations. A natural answer to this problem is to time-average response, but then it is not clear any more if a quick decision can be made. In the adaptive sorting mechanism, if we assume that $C_1$ is activating a downstream slow output, it can be shown that indeed, intrinsic fluctuations can be averaged out within tens of seconds, which is then compatible with the experimentally observed immune decision time and the appropriate sensitivity and specificity of the TCR signalling pathway~\cite{Lalanne:2013}.

\subsection{Antagonism}

In the previous section, we have considered  discrimination between ligands with different binding times when only one type of ligands is presented. However, a more realistic immunological situation is that agonist ligands are presented simultaneously with many sub threshold (self) ligands. So cells not only need to discriminate agonists from self ligands, but should also detect agonists presented within many self ligands.

It turns out that the adaptive sorting scheme presented in Figure~\ref{figure:AS}~A does not perform well:  addition of few sub thresholds ligands considerably decreases output concentration for the same agonist concentration. The fundamental reason is due to the coupling of $K$ between multiple receptors discussed in previous section.  Imagine self  ligands are added, calling $D$s corresponding complex (quality $\tau_s$), then we have for the phosphorylation activity:
\begin{equation}
\phi_K(C_0+D_0)=\frac{\phi}{C_0+D_0+C^*} \label{K_anta}
\end{equation}

and the total output concentration  is

\begin{equation}
C_1+D_1=\phi_K (C_0+D_0) (\tau C_0+\tau_s D_0)  \label{C_anta}
\end{equation}
For $\tau=\tau_s$, we naturally recover the same result as before, but if $\tau_s<<\tau$, the total output concentration clearly is much lower than the asymptotic response $\phi\tau$, an effect called antagonism. Intuitively, the ``self" ligand titrates the kinase necessary for the proofreading step, akin to a ``dog in the manger" effect described in other immune contexts \cite{Torigoe:1998vj}. If response is due to a thresholding effect, as a consequence, many more agonists ligands would be required to trigger response. In terms of response line of Figure~\ref{figure:absolute}~B, while specificity is conserved, the system loses sensitivity to minute concentrations of ligands.

%
%
%
%
%

It can be shown mathematically  that antagonism is a necessary consequence of absolute discrimination, qualifying as a ``phenotypic spandrel'' \cite{Francois:2016}. Intuitively, absolute discrimination necessariy requires some internal variable (similar to kinase $K$) to discriminate between ligands with different $\tau$s irrespective of their concentration, and as a consequence antagonism will always occur when those internal variables are activated by subthreshold ligands. A  model performing perfect absolute discrimination is presented in \cite{Francois:2016}, as well as a simple categorization of antagonistic effects, that can be increased or mitigated as $\tau\rightarrow 0$ depending on the model considered.

In particular there is a simple way to minimize the range of binding times with strong antagonism in the model of Figure~\ref{figure:AS}~A, by the addition of  a short upstream proofreading cascade as  discussed in ~\cite{Lalanne:2013}. Assuming now that kinase $K$ is activated after $m$ proofreading steps, we have 
\begin{equation}
\phi_K(C_m+D_m)=\frac{\phi}{C_m+D_m+C^*} 
\end{equation}
 But we have $C_m\propto \tau^{m+1}$ and $D_m \propto \tau_s^{m+1}$ so that $C_m >>D_m$ if $m$ is high enough, even if many self ligands are presented. As a consequence kinase $K$ is barely influenced by complex $D_m$. As said before, there nevertheless still is some antagonism close to threshold when $\tau_s \sim \tau$ which actually constitutes a smoking gun for absolute discrimination mechanisms (see next section for experimental evidence and \cite{Francois:2016} for a more careful study) . Antagonism is illustrated on Figure~\ref{figure:AS} B for a model with $4$ proofreading steps (response line is displayed on Figure~\ref{figure:absolute} D).

While antagonism is reduced by addition of proofreading steps, there are however other trade-offs appearing in the system: for instance adding too many proofreading steps might reduce the final concentration of the output too much, which creates several downstream problems in terms of response times very similar to what happens in McKeithan's KPR model (see Supplement of~\cite{Lalanne:2015} for discussions of this effect).

\begin{figure}
    \begin{center}
\resizebox{\textwidth}{!}{\includegraphics{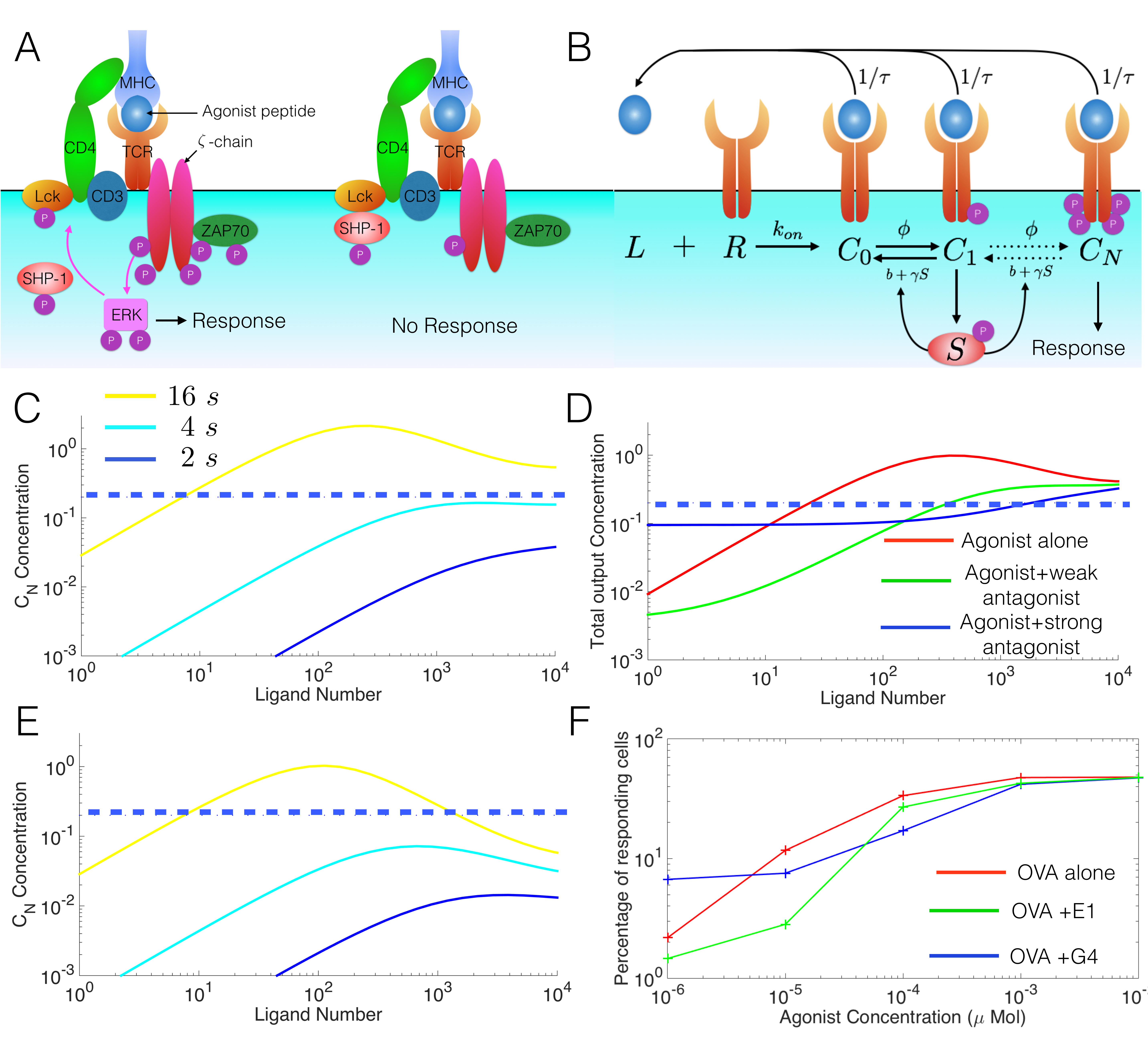}}
    \end{center}
    \caption{A simplified model for immune detection A: Sketch of some of the molecular players and their interactions. B: Simplified model based on kinetic proofreading combined with negative feedback C: $C_N$ concentration as a function of ligand for different values of $\tau$ D: Total output concentration as a function of agonist ($\tau=10 \ s$) ligand in presence of $10^4$ sub thresholds ligands, showing antagonism. E: behaviour of the network when $S$ total concentration is doubled, showing collapse at high ligand concentration. F: Experimental quantification of antagonism, reproduced from~\cite{Francois:2013}. OVA is agonist, E1 is weak antagonist, G4 is strong antagonist. } \label{figure:model} 
\end{figure}

\section{Biologically realistic models for immune detection}

As we have seen in previous section, generic solutions to the problem of absolute discrimination are now available within the simple adaptive sorting framework.  But one naturally wonders if biochemical reactions in actual T cells correspond to any of such Platonician view, and, if not, if they can be related to it in any ways. 

Adaptive sorting elaborates on a small number of kinetic proofreading steps. It should thus be first pointed out that many (if not all) molecular components of the original McKeithan model~\cite{Mckeithan:1995} are indeed present in actual cells. Such ``realistic" model explains its popularity and its use as a blueprint for many current models of immune decision \cite{Lever:2014}. The phosphorylation cascade would correspond to the known phosphorylation of Internal Tyrosine Activation Motifs (a.k.a. ITAM) containing chains in the TCR complex ~\cite{Kersh:1998a}. Rapid dephosphorylation upon unbinding would fit the kinetic segregation mechanism, specifying that generic phosphatases are segregated only upon ligand-receptor interaction~\cite{Davis:2006} . 

\subsection{Negative feedback and antagonism}

Adaptive sorting based models are expected to include an extra negative feedback, {\it i)} buffering for ligand concentration to realize adaptation/absolute discrimination, and  {\it ii)} associated with ligand antagonism.  This has been indeed observed and characterized in seminal papers by Dittel \etal~\cite{Dittel:1999}, and Stefanova \etal~\cite{Stefanova:2003}. These papers establish the existence of a negative component in ligand detection via the Tyrosine-protein phosphatase non-receptor type 6 (PTPN6), also known as Src homology region 2 domain-containing phosphatase-1 (SHP-1), and its role in ligand antagonism. In details, Dittel \etal quantified antagonism by measuring T cells responding to agonists in conjunction with increasing concentrations of antagonist ligands (which decreases the magnitude of immune response) in T cells endowed with two separate TCRs. Their measurement also suggested that antagonism is associated with SHP-1 association with TCR and the ensued dephosphorylation of ITAMs. Importantly, the antagonistic effect is ``infectious" within the cells: recruitment by SHP-1 spreads from receptor bound to antagonistic ligands even to unbound receptors, suggesting a global coupling of receptors via SHP-1. Involvement of SHP-1 in antagonism is definitely established in~\cite{Stefanova:2003}. They show in particular how moderate increase of SHP-1 activity gives several orders of magnitude increase of antagonism potency.
Antagonistic ligands are also shown to recruit more rapidly SHP-1 compared to agonist ligands (which only recruit SHP-1 in a later time). They finally show that agonist ligand specifically trigger a positive feedback loop, mediated by ERK-1, and that ERK-1 specifically inhibit SHP-1 recruitment by the TCR, thus explaining the kinetic difference between ligands. 

\subsection{Modeling negative feedback, approximate adaptive sorting}

The first model combining these different aspects was published by one of us (G.A--B) in collaboration with R.N Germain in 2005~\cite{AltanBonnet:2005}. This work combined a kinetic proofreading backbone with a SHP-1 mediated feedback and an ERK-1 positive feedback. We included most known components of the system, including different co-receptors, kinases, and eventual phosphorylation cascade in a very complex mathematical framework including around 300 dynamical variables. This model succeeded in satisfying the previously described golden triangle as a modeling target. It also establishes a clear linear hierarchy for antagonism, where stronger antagonists are ligands with binding times just below critical threshold $\tau_c$. Most importantly, the full-blown model was tested with new experiments quantifying precisely antagonism strength and decision time of the network, and validating predictions from the model.

While the Altan-Bonnet/Germain paper could explain many experimental features, it was not clear at that time if the full complexity of the model was required to understand the system. Can we see the ``biological wood emerg[ing] from the molecular tree" as nicely formulated by Gunawardena ~\cite{Gunawardena:2013}?  A first simplified model was proposed in 2008 by Lipniacki \etal~\cite{Lipniacki:2008}, but still contained around 40 variables, and was especially focusing on possible bistable properties of the system via ERK positive feedback loop. In 2013, we published in collaboration with G. Voisinne, E.D. Siggia \& M. Vergassola a considerably simplified version of this model~\cite{Francois:2013}, focusing on the part of the decision network upstream of ERK. Schematic of the model is displayed in Figure~\ref{figure:model}~B. Equations for this model are

\begin{eqnarray}
\label{eq1}
\dot{S} &=& \alpha C_1\left(S_T-S\right)-\beta S \label{S}\\
\label{eq2}
\dot{C_0}&=&\kappa (L_1-\sum_{i=0}^N C_i)(R-\sum_{i=0}^N C_i) \nonumber \\&&+(b+\gamma S)C_1-(\phi+\tau^{-1})C_0\\
\label{eq3}
\dot C_j&=&\phi C_{j-1}+(b+\gamma S)C_{j+1}-(\phi+b+\gamma S+\tau^{-1})C_j\\
\label{eq4}
\dot C_N&=&\phi C_{N-1}-(b+\gamma S+\tau^{-1})C_N
\end{eqnarray}

Like many other models, the backbone of this coarse-grained model relied on a kinetic proofreading backbone. Then, a simple negative feedback is included, in the form of the activation of a global phosphatase called $S$ (corresponding biologically to SHP-1) by a single complex in the phosphorylation cascade.

This model essentially recapitulates all observations made in 2005~\cite{AltanBonnet:2005},  satisfies the golden triangle, and is semi-analytic as detailed in the Supplement of \cite{Francois:2013}. In details, linearity of equations \ref{eq2} to \ref{eq4} in the limit of unsaturated receptors allows to easily express concentrations of all complex $C_j$ as a function of $S$ concentration. Defining  $r_+>r_-$ the roots of the characteristic equation that corresponds to equation \ref{eq3}

\begin{equation}
0 =\phi+(b+S)r^2-(\phi+b+S+\nu_1)r \label{root}
\end{equation}
we can show that 
\begin{equation}
C_j\simeq (1-r_-) \frac{\kappa R L \tau}{\kappa R \tau + 1}. r_-^j \label{cj}
\end{equation}

for $0<j<N$ and :

\begin{equation}
C_N\simeq \frac{\kappa R L \tau}{\kappa R \tau + 1}\left(1-\frac{r_-}{r_+}\right)r_-^{N}\label{CN:anal}
\end{equation}

One can then use equation \ref{S}  that relates $S$ to $C_1$ to close the system, which gives a fourth order polynomial equation that can be most easily solved numerically.

Response line of the model is illustrated on Figure~\ref{figure:absolute}D (``Immune model"). Interestingly, this network actually performs an approximate adaptive sorting: it flattens out the concentration of the output $C_N$ over several orders of magnitude of ligand concentration $L$, on different plateaus as a function of binding time $\tau$ as shown on Figure~\ref{figure:model}~C . Thus adaptive sorting appears to be the core principle of early immune detection, as could have been guessed in retrospect from first principles constraints evolved {\it in silico}. It is known that adaptation can be performed either via feedback or feedforward interactions, so that the network from Figure~\ref{figure:model}~B~can be seen as a feedback version of the feedforward adaptive sorting network presented in Figure~\ref{figure:AS} .  Another difference with adaptive sorting as discussed before is that the same kinase and phosphatase is shared  between all proofreading steps, while networks such as the ones displayed on Figure~\ref{figure:AS} require one specific kinase for the step actually performing adaptation. Our model implements approximate adaptive sorting with a minimum set of shared unspecific kinases and phosphatases, and as such can be considered as optimum in terms of parsimony of biochemical species used. Full stochastic simulations show that decision times are consistent with data from \cite{AltanBonnet:2005}  and can be as low as $30$ s for strong agonists. Finally, only adaptive sorting models with several proofreading steps such as the one of Figure~\ref{figure:AS}~B and Figure \ref{figure:model} give antagonistic properties similar to experimental data Figure~\ref{figure:model}~F~\cite{Francois:2013}.

Intuititive analytical limits can be obtained when negative feedback does not saturate, which is biologically relevant given the observed phenotypic variability (see \cite{Feinerman:2008a} and next section). For instance, if we assume that phosphatase $S$ (corresponding to SHP-1) is in excess and that $b<<1$ , in the limit of high ligand, we can show that $C_1 \propto S \propto \sqrt{L}$ using equations \ref{S}, \ref{root} and \ref{cj}, and characteristic equation \ref{root} gives $r_+\sim 1$ and $r_-\propto \phi/S$. The latter expression of $r_-$ is particularly interesting, being a ratio of the forward phosphorylation rate $\phi$ over the backward dephosphorylation rate scaling like $S$. From equation \ref{CN:anal}, it appears that the balance of these two rates essentially determines level of response, which fits the intuitive biological idea that the main role of the negative feedback is to restrain progress in the proofreading cascade.

Putting everything together, for high concentration of agonist ligands, irrespective of binding times, a simple asymptotic relationship holds

\begin{equation}
C_N\propto L^{1-\frac{N}{2}}
\end{equation}
  connecting ligand concentration $L$ and $C_N$, with $N$  the number of proofreading steps. So if $N$ is high enough, $C_N$ decreases with ligand concentration $L$ in this model. This seems rather counter-intuitive, and predicts that if feedback is strong enough, immune response at high agonist ligand concentration could thus disappear (see Figure \ref{figure:model} ~E, full stochastic treatment can be found in \cite{Francois:2013} ). Indeed, we tested and verified this prediction in cells with high level of SHP-1 \cite{Francois:2013}. Furthermore, if SHP-1 level is increased above a couple a few-fold, negative feedback essentially dominates for all ligand concentration, and response is fully abolished, a ``digital effect" indeed experimentally observed in~\cite{Feinerman:2008a} and first predicted with  the Altan-Bonnet Germain model \cite{AltanBonnet:2005}.

\subsection{Beyond absolute discrimination ?}

Other modelling strategies for immune recognition by single cells have been proposed. Major differences in most models is that  $\tau$ dependency is not considered as sharp as what is assumed here, and modulation by other parameters is considered.  For instance it has been suggested that there might actually be an optimal time of interaction (i.e. response would decrease at higher $\tau$s) \cite{Kalergis:2001fn}. This effect is debated, since strong binding TCRs evolved {\it in vitro}, with $\tau$s of the order of $50$ to $100$ s conversely yield an exceptionally potent response  \cite{Donermeyer:2006hi}. Lever \etal recently reviewed and compared 5 families of models \cite{Lever:2014}, including the one from previous section, and suggested that a mid-range $\tau$ can be simply explained by a kinetic proofreading with ``limited' signalling'', meaning that the signalling complex in the cascade gets inactivated with constant rate. However, this model does not include the negative feedback described here, and thus do not display antagonism. Another issue of interest is the regime of fast ``on rates''': for such ligands, response is experimentally observed even for low $\tau$s. It has been suggested that  competion between binding/rebinding and diffusion of receptors on the membrane would actually give an effective higher $\tau$ \cite{Govern:2010kx}, which therefore would not necessarily contradict the life-time dogma but rather complexify it.  A recent review \cite{Chakraborty:2014hw} summarizes  molecular players of the systems and relationships to proofreading mechanisms, discussing in particular the role of co-receptors and various kinases, and possible other mechanisms that could explain better high sensitivity such as serial triggering by one ligand.

Other theoretical models have explored biophysical limits or more refined computations. For instance, a related question is the minimum theoretical decision time for such process ( ``decision on the fly"). As said in introduction, a trade-off between accuracy and precision is expected, and this is a potentially acute issue in an immune context. If decision takes too much time, a foreign ligand might be gone before proper response is activated. If accuracy is decreased, there is potential for auto-immune response. Exact results for this problem have been recently obtained  by Siggia and Vergassola ~\cite{Siggia:2013}. They study the problem of detection of change of composition of ligand mixtures. It turns out that Wald sequential probability test ratio on the log likelihood of a sequence of binding events can be used to take decision. Strikingly,  simple phosphorylation networks reminiscent of network controlling early immune detection can naturally implement biological versions of this test~\cite{Siggia:2013}. An important aspect to optimize decision time is that decision here is made ``dynamically", while   networks performing decision here essentially work at steady state (see nevertheless Supplement of \cite{Lalanne:2013} for discussions of transient behaviours).  Such dynamical sensing could also help in the regime of very small ligand concentrations, maybe in conjunction with serial triggering \cite{Chakraborty:2014hw}. An even more general problem is chemodetection of ligands of different qualities in fluctuating environments, where mixture composition can vary wit time. If there is a high probability of observing ligands just subthresholds, antagonism is helpful to essentially buffer spurious fluctuations of ligands~\cite{Lalanne:2015}, reminiscent of what is observed in olfaction \cite{Tsitron:2011hq}. Finally, some models have recently explored theoretically the possible network topologies maximizing information transmission on ligand mixtures, and have pointed out the need of both proofreading and internal feedbacks \cite{Mora:2015cv,Singh:2015vs}.

\section{Ligand recognition by a population of lymphocytes: more is different?}

\subsection{Tackling cell-to-cell variability in immune responses: statistical physics to the rescue?}

There are still many challenges to fully understand early immune response. In particular, once a cell has been activated, it appears that further processing occurs at the immune population level. Modeling the early events in immune detection potentially is very topical for statistical physics as it involves the accounting of cell-to-cell variabilities, modeling immune responses with distribution of cells, and the testing of their functional significance.

Starting with the seminal work of M. Elowitz \etal ~\cite{Elowitz:2002}  a subfield of biological physics has grown to address the emergence of cell-to-cell variability\footnote{note that, although physicists are fond to describe such variability as ``noise'', based on their representation using a Langevin equation, this term remains confusing for most biologists because of its negative connotation: we elicit to use the more neutral term of cell-to-cell variability}. Extrinsic variability  stems from many sources (e.g. all sources that do not relate to the stochasticity and low-copy number in biochemical reactions): they can include heterogeneity in the cellular environment, as well as epigenetic variation and metabolic fluctuations from cell-to-cell. Intrinsic contributions  can be best studied  in an isogenic population of cells and are the physical consequences of stochasticity in the biochemical reactions. As a consequence, cells can display broad distributions of phenotypes based on the heterogeneity of abundance of key regulatory proteins (receptors, kinases and phosphatases in the context of signal transduction network; transcription factors in the context of gene regulation).

First forays to address the functional significance of cell-to-cell variability in biological systems were focused on bacterial responses. From chemotaxis (the ability to orient motions in gradients of nutrients or chemokines) to competence the ability to acquire new genomic materials), researchers demonstrated that, indeed, varied levels of key proteins could map into varied phenotypes~\cite{Korobkova:2004,CaGatay:2009}. Such observations were used at first as new quantitative constraints to validate biochemical models of  biological regulation. 

Concomitantly, these observations demonstrated that a ``mean-field" measurement and model of a population of cells might have serious shortcomings when predicting global responses. One example where such cell-to-cell variability was found to be critical is in the study of bacterial antibiotic resistance. For example, Balaban and coworkers introduced a microfluidic device to track the proliferation and death of bacteria under antibiotic treatment~\cite{Balaban:2004,Rotem:2010}. A sub population of isogenic bacteria were found to resist to antibiotics, simply by being a different metabolic state compared to their sister cells at the time of exposure to antibiotics. Such process of distributed response based on distributed phenotype at stimulation time was described as an optimal strategy to tune responses in a fluctuating environment, by matching probability of phenotypic switching to probability of environmental changes ~\cite{Kussell:2005}. Indeed, there is such a fundamental mismatch between the necessary response time (cells must exclude antibiotics on very short timescales) and the evolutionary constraints (it would take a large amount of time and a very low probability for cells to generate a solution to the problem of antibiotics resistance), that cells are better off diversifying their phenotype pre-emptively such that a solution is readily accessible when the antibiotic perturbation applies. 

\subsection{Phenotypic variability of T cell ligand discrimination.}
\label{sssec:variability}

Similar constraints are at play in the context of the immune response. There is a similar disconnect between the dynamics of biological problems at stake (eradicating a fast-replicating fast evolving pathogen vs. generating an adaptive immune response). In particular, recent measurements by the Jenkins \& Davis lab \cite{Nelson:2015kc,Han:2014jl} have beautifully illustrated the number of constraints for a good immune response. Lymphocytes can rapidly proliferate (by factors of $10^3$ to $10^6$) and relax back to low numbers for the memory pool. Such explosive proliferation is critical to match the challenge posed by fast-proliferating pathogens. However, surprisingly,  the number of T cell clones that can recognize a specific antigen (e.g. flu peptide) is very small, with 10 to 100 clones per individual (mouse or human).  This is particularly relevant as pathogens can be ``cunning" and attempting to invoke multiple escape mechanisms. Hence, rather than a deterministic adaptation of the immune response to the pathogenic challenge and optimization of detection at the cellular scale, one can conjecture that the immune system relies on some degree of statistical randomness in order to diversify responses within an isogenic population of lymphocytes.

The functional underpinnings of such cell-to-cell variability can be readily detected in the context of the early events of immune detection. Indeed, researchers have been using the vast panoply of antibodies to quantify protein and phospho-protein levels in cells, as well as the single-cell resolution of cytometry (using fluorescence-based or mass-spectrometry-based approaches), in order to quantify the cell-to-cell variability of responses to external stimuli. In a nutshell, if the abundance of protein X is limiting in the activation of Y into Y* (e.g. phosphorylation), then measuring and correlating X with Y* abundances at the single cell level will reveal heterogeneity of the response. Such cell-to-cell variability analysis has been carried out in the context of immune detection to demonstrate the sensitivity of ligand discrimination to varied levels of signaling components (e.g. CD8 and SHP-1)~\cite{Feinerman:2008a}, as well as the sensitivity of T cells response to the cytokine IL-2 to varied levels of cytokine receptors (e.g. CD25, CD122, and CD132 a.k.a. IL-2R$\alpha$, IL-2R$\beta$ and 
$\gamma_{C}$)~\cite{Cotari:2013}. Note that such parameter sensitivities were first predicted from the dynamical model of the signaling cascades at play,  and Cell-to-Cell Variability Analysis (CCVA, a new methodology introduced in~\cite{Cotari:2013}) validated these predictions quantitatively. Similarly, single-cell analysis has been carried out in the space of phospho-proteins by the Pe'er \& Nolan labs to quantify the strength of connections within the TCR signaling pathway~\cite{Krishnaswamy:2014}. There, analysis of single-cell measurements using overall, resolution of immune responses at the individual cell level has highlighted the large phenotypic variability in the signaling response of individual lymphocytes: such observations must then be interpreted functionally to map out how T cells diversify their response to optimize its detection capabilities. 

\subsection{Deriving reliable immune responses from unreliable responses of individual cells.}
\label{section:cooptation}

The cell-to-cell variability of lymphocyte response presents a challenge for our current cell-centric understanding of self/not-self discrimination in the immune system. Indeed, if each individual T cell makes a rapid, sharp yet utterly variable decision to respond to a ligand (see~\ref{sssec:variability}), one could anticipate many ``mistakes" whereby T cells would respond to self tissues and trigger an auto--immune disorder. 

We conjecture that it is the integration of the responses of individual T cells over longer timescales ($> hour$) that may correct for the ``sloppiness" of individual T cell on short timescales ($\approx$ minutes). Such integration can be carried out through cell-cell communications {\it e.g.} through secretion and consumption of cytokines. In particular, K. Tkach {\it et al.} measured experimentally how much cytokine accumulates in the supernatant of T cells that were activated {\it in vitro}~\cite{Tkach:2014}. A surprising scaling law for this $IL-2_{max}$ capacity with the regards to the quantity of antigens $\#(pMHC)$ and the size of the T cell population is observed
\begin{equation}
[IL\mhyphen2]_{max} \propto \left(\#(pMHC)\right)^{0.8}.\left(N_{T cells}\right)^{-0.1}
\end{equation}

Hence, T cells are able to output cytokines in a near-linear scalable manner across four decades of $\#(pMHC)$: this is quite a striking observation when considering the limited dynamic range of individual T cell response and more generally of any biological system. Moreover, the independence of output with the number of T cells $N_{T cells}$ in the system is remarkable considering that cytokine secretion is {\it a priori} an extensive variable: this finding implies that the overall capacity of the system (in terms of maximal cytokine concentration) is an intensive variable. Deriving an intensive output ({\it i.e.} an output that is independent from the size of the system) for a population of T cells has been observed in very related experiments~\cite{Hart:2014}.  Hart {\it et al.} measured the cell expansion of a population of CD4$^+$ T cells after activation through their antigen receptor pathway. Hart {\it et al.}'s model focused on the cytokine IL-2 whose divergent function (cell proliferation and apoptosis) would explain such robust system-size-independent output. 

Understanding mechanistically the emergence of such scaling laws implied revisiting our classical biochemical understanding of the IL-2 cytokine pathway. It was well established that T cells would essentially shut down their IL-2 production as soon as they secreted and built up a pool of shared cytokine. 
 Biochemically, this implied that the IL-2 output of a population of T cells would have a ceiling of $10\ pM$, as the concentration of cytokine required to induce IL-2 signaling and subsequent shutdown to IL-2 secretion~\cite{Waysbort:2013}. However, detailed quantitation of the biochemistry of the system unraveled a negative cross-talk from the antigen signaling response to the IL-2 response pathway~\cite{Tkach:2014}. When measuring the phosphorylation of the $STAT5$ transcription factor downstream of IL-2 sensing, Tkach {\it et al.} uncovered a surprising convolution between response to $pMHC$ antigens and $IL\mhyphen2$ cytokine:
\begin{equation}
\#(pSTAT5)\propto \frac{\#(IL\mhyphen2R\beta/\gamma_C)}{1+\frac{K_D}{\#(IL\mhyphen2R\alpha).[IL-2]}}.\frac{1}{1+\frac{\#(pMHC)}{\#(pMHC)_0}}
\end{equation}

\noindent where $\#(X)$ represents the number of $X$ within the cell, $\#(IL\mhyphen2R\alpha)$ is the number of $\alpha$ chain of the $IL-2$ receptor,  and $\#(IL\mhyphen2R\beta/\gamma_C)$ is the number of complexed  $IL\mhyphen2R\beta$ and $\gamma_C$ receptors on the surface of T cells --in a nutshell, $\#(IL\mhyphen2R\alpha)$ gets expressed in activated T cells to complete the signaling $\#(IL\mhyphen2R\beta/\gamma_C)$ pair into a signaling complex, upon $IL\mhyphen2$ binding~\cite{Cotari:2013}. Overall, this equation encapsulates the  convolution between local and global responses ({\it i.e.} $pMHC$ abundances  and $IL\mhyphen2$ concentration respectively): this convolution is particularly significant as it regulates the {\it off} switch for $IL\mhyphen2$ production. This is also important because $pSTAT5$ in turns regulate IL-2, which means that there is a feedback between the local stimulation (by $pMHC$ ) and the global readout (by $IL\mhyphen2$) through the regulation of $STAT5$ phosphorylation. Overall, a combination of an incoherent feedforward loop  was shown to be necessary and sufficient to explain the scaling law in the accumulation of $IL\mhyphen2$ in the milieu. 

Such initial attempt by Tkach \etal~\cite{Tkach:2014} was based on detailed biochemical modeling with explicit biochemistry being implemented. Additional experiments revealed additional functional relevance for the secretion of IL-2, as a global regulator of self/not-self discrimination: Voisinne {\it et al.} probed the proliferation response of a population of T cells containing multiple clones of diverse specificity ({\it i.e.} a polyclonal population of cells). This study demonstrated that strongly activated T cell clones could induce the activation and proliferation of neighboring weakly activated T cell clones ~\cite{Voisinne:2015}: this previously--undocumented phenomenon of propagating T cell activation was termed ``T cell co-optation''. Similarly to previous studies~\cite{AltanBonnet:2005,Feinerman:2010,Tkach:2014}, an experimentally-derived biochemically-explicit model of such multi scale integration was introduced to account for such lymphocyte co-optation. Overall, it was demonstrated that the addition of antigen signals (read locally through the TCR pathway) and cytokine signals (read globally through the IL-2 pathway) decides T cell fate. Such observations expand over longer timescales (days) and larger timescales (lymph nodes) what individual T cell can contribute in terms of immune response.  This is particularly significant as it demonstrated that T cell activation can no longer be considered as a local property of antigen recognition, but gets decided by integrating multiple cues. The practical implication should not be understated when manipulating the response of a polyclonal population of T cells to force weak T cell clones to respond to antigens ({\it e.g.} in the context of tumors) may hold the key to tumor eradication. Future efforts will require phenomenological coarse-graining to allow better understanding of immune recognition at the level of the system.

\begin{figure}
\label{figure:CollectiveRecognition} 

    \begin{center}
\resizebox{\textwidth}{!}{\includegraphics{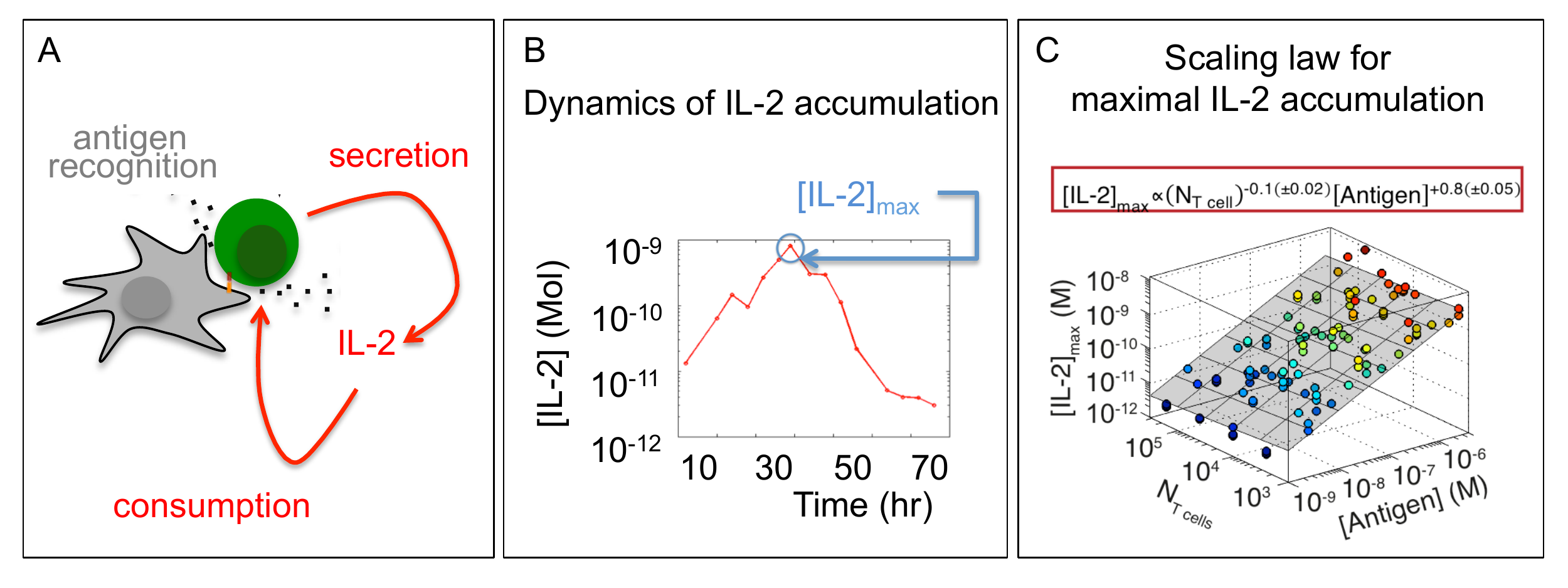}}
    \end{center}
    \caption{Immune recognition over long timescales may involve collective synchronization of T cell activation. While T cells perform an absolute (immune) discrimination at the individual cell level, their heterogenous expression of signaling components leads to unreliability in their threshold of activation. Moreover, experimental work on lymphocyte signaling demonstrated that individual T cells respond in an all-or-none (digital) such that their output contains limited information.  Additional mechanisms must be at play to ``proofread" individual immune recognition and/or generate plastic immune responses. Feedback regulations based on cytokine secretion and consumption constitute an efficient solution to the limitations of individual T cells~\cite{Feinerman:2010,Hart:2014,Tkach:2014,Youk:2014}. In particular, T cells can rely on such cell-to-cell communications to achieve higher levels of immune recognition. {\bf B.}  Tkach {\it et al.} found a surprising scaling law, whereby the maximum concentration ($[IL-2]_{max}$) of the IL-2 cytokine released by a population of T cells scales almost linearly with the amount of antigens that is present in the system, practically independently of the number of T cells present in the system. {\bf C.} Such analog scaling at the population level was found to derive through a coherent feed-forward loop of Type 4, using the nomenclature introduced by U. Alon and coworkers~\cite{Mangan:2003}.}
\end{figure}

Recent technical developments to monitor T cell responses at the individual cell level are enabling researchers to track the early events of immune activation, one cell at a time. Monitoring the differentiation of individual lymphocytes will certainly accelerate our study of the immune system, in particular when stochastic effects and phenotypic variability are necessary to explain the diversification in immune detection. Yet, in the context of the study of the immune system as a whole, it is the collective properties of the cells based on their cytokine communications and competition for antigen that shape the overall immune response.

Finally, on the pure theory side, some recent questions inspired by statistical physics include how the immune ligand landscape is matched by receptors/cellular diversity, and how collective decisions can be made within this framework, without entering into specific signalling details. A simplified probabilistic model of T-cell activation has been derived using extreme values theories (modelling  binding energy of TCRs to single ligand), from which probability of false positive can be computed\cite{Butler:2013hc}, and collective decision was shown to improve with the help of cell to cell communications. Mayer \etal very recently proposed a general framework, predicting that more receptors are needed for rate antigens, and strong cross-reactivity to correctly cover the full antigen landscape \cite{Mayer:2015ce}. 
The future will certainly involve more statistical-physics based approaches that tackle the large number of lymphocytes and focus on such emergent properties.


\section{Wrapping up.}

To conclude, we presented an overview of some physics-inspired studies in immunology, focused on early recognition by T-Cells. Specifically, we discussed how the core problem of the adaptive immune response (the recognition of pathogen-derived antigens) must be addressed with quantitative models. Indeed, experimental results have now well established the biophysical underpinnings of self/not-self discrimination namely the fact that small increases in the lifetime of ligand-receptor complexes lead to large increases in ligand potency. Additionally, cells can respond very sensitively and with great speed. Quantitatively reconciling these three experimental aspects (so-called golden triangle) is a theoretical challenge with great conceptual and practical relevance (e.g. when fine-tuning T cell activation is required, as in cancer immunotherapies).
	We reviewed the current states of theoretical models, building upon the original kinetic proofreading scheme. {\it In silico} evolution of biochemical networks satisfying the golden triangle unraveled a minimal model that can reconcile all aspects of ligand discrimination. In particular, a proximal negative feedback (associated with the activation of a phosphatase) was found to be critically relevant to abrogate responses to self ligands (even in large quantities) while allowing responses to not-self ligands (even in small quantities). The functional pay-off of these models is to ``predict" the existence of antagonism in immune recognition as well.
	Finally, we discussed how these models of ligand discrimination by T cells create new challenges in terms of understanding the phenotypic variability of isogenic populations of T cells, or in terms of accounting for the quantitative response to antigens when measuring T cell activation over long timescales. Similar collaborations between experimentalists and theoretical physicists will remain fruitful to expand our quantitative understanding of T cell activation to more complex issues in immunology (role of regulatory T cells, tuning of responsiveness according to inflammatory milieu etc.). More generally, we hope that these fundamental issues of immunology will spark the interest of statistical physicists, as the derivation and manipulation of large-scale immune response from the local activation of individual T cells remains poorly understood at the theoretical level.


\begin{thebibliography}{10}
\providecommand{\url}[1]{{#1}}
\providecommand{\urlprefix}{URL }
\expandafter\ifx\csname urlstyle\endcsname\relax
  \providecommand{\doi}[1]{DOI~\discretionary{}{}{}#1}\else
  \providecommand{\doi}{DOI~\discretionary{}{}{}\begingroup
  \urlstyle{rm}\Url}\fi

\bibitem{Allard:2012}
Allard, J.F., Dushek, O., Coombs, D., van~der Merwe, P.A.: {Mechanical
  Modulation of Receptor-Ligand Interactions at Cell-Cell Interfaces}.
\newblock Biophysical Journal \textbf{102}(6), 1265--1273 (2012)

\bibitem{AltanBonnet:2005}
Altan-Bonnet, G., Germain, R.N.: {Modeling T Cell Antigen Discrimination Based
  on Feedback Control of Digital ERK Responses}.
\newblock PLoS Biology \textbf{3}(11), e356 (2005)

\bibitem{Artomov:2007}
Artyomov, M.N., Das, J., Kardar, M., Chakraborty, A.K.: {Purely stochastic
  binary decisions in cell signaling models without underlying deterministic
  bistabilities}.
\newblock Proc Natl Acad Sci U S A \textbf{104}(48), 18,958--18,963 (2007)

\bibitem{Balaban:2004}
Balaban, N.Q., Merrin, J., Chait, R., Kowalik, L., Leibler, S.: {Bacterial
  persistence as a phenotypic switch}.
\newblock Science \textbf{305}(5690), 1622--1625 (2004)

\bibitem{Bialek:2012}
Bialek, W.: {Biophysics: searching for principles}.
\newblock Princeton University Press (2012)

\bibitem{Block:1983}
Block, S.M., Segall, J.E., Berg, H.C.: {Adaptation kinetics in bacterial
  chemotaxis}.
\newblock Journal of bacteriology \textbf{154}(1), 312--323 (1983)

\bibitem{Butler:2013hc}
Butler, T.C., Kardar, M., Chakraborty, A.K.: {Quorum sensing allows T cells to
  discriminate between self and nonself}.
\newblock Proc Natl Acad Sci U S A \textbf{110}(29), 11,833--11,838 (2013)

\bibitem{CaGatay:2009}
CaGatay, T., Turcotte, M., Elowitz, M.B., Garcia-Ojalvo, J., S{\"u}el, G.M.:
  {Architecture-Dependent Noise Discriminates Functionally Analogous
  Differentiation Circuits}.
\newblock Cell \textbf{139}(3), 512--522 (2009)

\bibitem{Celani:2010}
Celani, A., Vergassola, M.: {Bacterial strategies for chemotaxis response}.
\newblock Proceedings of the National Academy of Sciences of the United States
  of America \textbf{107}(4), 1391--1396 (2010)

\bibitem{Chakraborty:2014hw}
Chakraborty, A.K., Weiss, A.: {Insights into the initiation of TCR signaling}.
\newblock Nature immunology \textbf{15}(9), 798--807 (2014)

\bibitem{Corson:2012}
Corson, F., Siggia, E.D.: {Geometry, epistasis, and developmental patterning.}
\newblock Proc Natl Acad Sci U S A \textbf{109}(15), 5568--5575 (2012)

\bibitem{Cotari:2013}
Cotari, J.W., Voisinne, G., Dar, O.E., Karabacak, V., Altan-Bonnet, G.:
  {Cell-to-Cell Variability Analysis Dissects the Plasticity of Signaling of
  Common Chain Cytokines in T Cells}.
\newblock Science Signaling \textbf{6}(266), ra17--ra17 (2013)

\bibitem{Davis:2006}
Davis, S.J., van~der Merwe, P.A.: {The kinetic-segregation model: TCR
  triggering and beyond.}
\newblock Nature immunology \textbf{7}(8), 803--809 (2006)

\bibitem{Dittel:1999}
Dittel, B.N., Stefanova, I., Germain, R.N., Janeway, C.A.: {Cross-antagonism of
  a T cell clone expressing two distinct T cell receptors}.
\newblock Immunity \textbf{11}(3), 289--298 (1999)

\bibitem{Donermeyer:2006hi}
Donermeyer, D.L., Weber, K.S., Kranz, D.M., Allen, P.M.: {The Study of
  High-Affinity TCRs Reveals Duality in T Cell Recognition of Antigen:
  Specificity and Degeneracy}.
\newblock The Journal of Immunology \textbf{177}(10), 6911--6919 (2006)

\bibitem{Dushek:2011}
Dushek, O., Aleksic, M., Wheeler, R.J., Zhang, H., Cordoba, S.P., Peng, Y.C.,
  Chen, J.L., Cerundolo, V., Dong, T., Coombs, D., van~der Merwe, P.A.:
  {Antigen Potency and Maximal Efficacy Reveal a Mechanism of Efficient T Cell
  Activation}.
\newblock Science Signaling \textbf{4}(176), ra39--ra39 (2011)

\bibitem{Elowitz:2002}
Elowitz, M.B., Levine, A.J., Siggia, E.D., Swain, P.S.: {Stochastic gene
  expression in a single cell.}
\newblock Science \textbf{297}(5584), 1183--1186 (2002)

\bibitem{Feinerman:2008b}
Feinerman, O., Germain, R.N., Altan-Bonnet, G.: {Quantitative challenges in
  understanding ligand discrimination by alphabeta T cells.}
\newblock Molecular immunology \textbf{45}(3), 619--631 (2008)

\bibitem{Feinerman:2010}
Feinerman, O., Jentsch, G., Tkach, K.E., Coward, J.W., Hathorn, M.M., Sneddon,
  M.W., Emonet, T., Smith, K.A., Altan-Bonnet, G.: {Single-cell quantification
  of IL-2 response by effector and regulatory T cells reveals critical
  plasticity in immune response}.
\newblock Molecular Systems Biology \textbf{6} (2010)

\bibitem{Feinerman:2008a}
Feinerman, O., Veiga, J., Dorfman, J.R., Germain, R.N., Altan-Bonnet, G.:
  {Variability and Robustness in T Cell Activation from Regulated Heterogeneity
  in Protein Levels}.
\newblock Science \textbf{321}(5892), 1081--1084 (2008)

\bibitem{Francois:2014}
Fran{\c c}ois, P.: {Evolving phenotypic networks in silico.}
\newblock Seminars in cell {\&} developmental biology \textbf{35}, 90--97
  (2014)

\bibitem{Francois:2016}
Fran{\c c}ois, P., Johnson, K.A., Saunders, L.N.: {Phenotypic spandrel:
  absolute discrimination and ligand antagonism}.
\newblock arXiv.org  (2015)

\bibitem{Francois:2013}
Fran{\c c}ois, P., Voisinne, G., Siggia, E.D., Altan-Bonnet, G., Vergassola,
  M.: {Phenotypic model for early T-cell activation displaying sensitivity,
  specificity, and antagonism.}
\newblock Proc Natl Acad Sci U S A pp. 1--15 (2013)

\bibitem{Gascoigne:2001}
Gascoigne, N.R., Zal, T., Alam, S.M.: {T-cell receptor binding kinetics in
  T-cell development and activation.}
\newblock Expert reviews in molecular medicine \textbf{2001}, 1--17 (2001)

\bibitem{Govern:2010kx}
Govern, C.C., Paczosa, M.K., Chakraborty, A.K., Huseby, E.S.: {Fast on-rates
  allow short dwell time ligands to activate T cells.}
\newblock Proc Natl Acad Sci U S A \textbf{107}(19), 8724--8729 (2010)

\bibitem{Gregor:2007}
Gregor, T., Tank, D.W., Wieschaus, E.F., Bialek, W.: {Probing the limits to
  positional information.}
\newblock Cell \textbf{130}(1), 153--164 (2007)

\bibitem{Gunawardena:2013}
Gunawardena, J.: {Models in biology: 'accurate descriptions of our pathetic
  thinking'.}
\newblock BMC biology \textbf{12}(1), 29--29 (2013)

\bibitem{Han:2014jl}
Han, A., Glanville, J., Hansmann, L., Davis, M.M.: {Linking T-cell receptor
  sequence to functional phenotype at the single-cell level}.
\newblock Nature biotechnology \textbf{32}(7), 684--692 (2014)

\bibitem{Hart:2014}
Hart, Y., Reich-Zeliger, S., Antebi, Y.E., Zaretsky, I., Mayo, A.E., Alon, U.,
  Friedman, N.: {Paradoxical signaling by a secreted molecule leads to
  homeostasis of cell levels.}
\newblock Cell \textbf{158}(5), 1022--1032 (2014)

\bibitem{Holler:2000}
Holler, P.D., Holman, P.O., Shusta, E.V., O'Herrin, S., Wittrup, K.D., Kranz,
  D.M.: {In vitro evolution of a T cell receptor with high affinity for
  peptide/MHC}.
\newblock Proc Natl Acad Sci U S A \textbf{97}(10), 5387--5392 (2000)

\bibitem{Hopfield:1974}
Hopfield, J.J.: {Kinetic proofreading: a new mechanism for reducing errors in
  biosynthetic processes requiring high specificity}.
\newblock Proceedings of the National Academy of Sciences of the United States
  of America \textbf{71}(10), 4135--4139 (1974)

\bibitem{Huang:2013}
Huang, J., Brameshuber, M., Zeng, X., Xie, J., Li, Q.J., Chien, Y.h.,
  Valitutti, S., Davis, M.M.: {A single peptide-major histocompatibility
  complex ligand triggers digital cytokine secretion in CD4(+) T cells.}
\newblock Immunity \textbf{39}(5), 846--857 (2013)

\bibitem{Huang:2010df}
Huang, J., Zarnitsyna, V.I., Liu, B., Edwards, L.J., Jiang, N., Evavold, B.D.,
  Zhu, C.: {The kinetics of two-dimensional TCR and pMHC interactions determine
  T-cell responsiveness.}
\newblock Nature \textbf{464}(7290), 932--936 (2010)

\bibitem{Kalergis:2001fn}
Kalergis, A.M., Boucheron, N., Doucey, M.A., Palmieri, E., Goyarts, E.C., Vegh,
  Z., Luescher, I.F., Nathenson, S.G.: {Efficient T cell activation requires an
  optimal dwell-time of interaction between the TCR and the pMHC complex}.
\newblock Nature immunology \textbf{2}(3), 229--234 (2001)

\bibitem{Kersh:1998a}
Kersh, E.N., Shaw, A.S., Allen, P.M.: {Fidelity of T cell activation through
  multistep T cell receptor zeta phosphorylation.}
\newblock Science \textbf{281}(5376), 572--575 (1998)

\bibitem{Kersh:1998wx}
Kersh, G.J., Kersh, E.N., Fremont, D.H., Allen, P.M.: {High- and low-potency
  ligands with similar affinities for the TCR: The importance of kinetics in
  TCR signaling}.
\newblock Immunity \textbf{9}(6), 817--826 (1998)

\bibitem{Korobkova:2004}
Korobkova, E., Emonet, T., Vilar, J.M.G., Shimizu, T.S., Cluzel, P.: {From
  molecular noise to behavioural variability in a single bacterium}.
\newblock Nat Cell Biol \textbf{428}(6982), 574--578 (2004)

\bibitem{Krishnaswamy:2014}
Krishnaswamy, S., Spitzer, M.H., Mingueneau, M., Bendall, S.C., Litvin, O.,
  Stone, E., Pe'er, D., Nolan, G.P.: {Systems biology. Conditional
  density-based analysis of T cell signaling in single-cell data.}
\newblock Science \textbf{346}(6213), 1250,689--1250,689 (2014)

\bibitem{Kussell:2005}
Kussell, E., Leibler, S.: {Phenotypic diversity, population growth, and
  information in fluctuating environments.}
\newblock Science \textbf{309}(5743), 2075--2078 (2005)

\bibitem{Lalanne:2013}
Lalanne, J.B., Fran{\c c}ois, P.: {Principles of adaptive sorting revealed by
  in silico evolution.}
\newblock Physical Review Letters \textbf{110}(21), 218,102 (2013)

\bibitem{Lalanne:2015}
Lalanne, J.B., Fran{\c c}ois, P.: {Chemodetection in fluctuating environments:
  Receptor coupling, buffering, and antagonism.}
\newblock Proc Natl Acad Sci U S A  (2015)

\bibitem{Lever:2014}
Lever, M., Maini, P.K., van~der Merwe, P.A., Dushek, O.: {Phenotypic models of
  T cell activation}.
\newblock Nature Reviews Immunology \textbf{14}(9), 619--629 (2014)

\bibitem{Lipniacki:2008}
Lipniacki, T., Hat, B., Faeder, J.R., Hlavacek, W.S.: {Stochastic effects and
  bistability in T cell receptor signaling}.
\newblock Journal of Theoretical Biology \textbf{254}(1), 110--122 (2008)

\bibitem{Liu:2014}
Liu, B., Chen, W., Evavold, B.D., Zhu, C.: {Accumulation of Dynamic Catch Bonds
  between TCR and Agonist Peptide-MHC Triggers T Cell Signaling}.
\newblock Cell \textbf{157}(2), 357--368 (2014)

\bibitem{Mangan:2003}
Mangan, S., Alon, U.: {Structure and function of the feed-forward loop network
  motif}.
\newblock Proceedings of the National Academy of Sciences of the United States
  of America \textbf{100}(21), 11,980--11,985 (2003)

\bibitem{Mayer:2015ce}
Mayer, A., Balasubramanian, V., Mora, T., Walczak, A.M.: {How a well-adapted
  immune system is organized.}
\newblock Proc Natl Acad Sci U S A \textbf{112}(19), 5950--5955 (2015)

\bibitem{Mckeithan:1995}
McKeithan, T.W.: {Kinetic Proofreading in T-Cell Receptor Signal-Transduction}.
\newblock Proceedings of the National Academy of Sciences of the United States
  of America \textbf{92}(11), 5042--5046 (1995)

\bibitem{Mehta:2012ji}
Mehta, P., Schwab, D.J.: {Energetic costs of cellular computation.}
\newblock Proc Natl Acad Sci U S A \textbf{109}(44), 17,978--17,982 (2012)

\bibitem{Mora:2015cv}
Mora, T.: {Physical Limit to Concentration Sensing Amid Spurious Ligands}.
\newblock Physical Review Letters \textbf{115}(3), 038,102 (2015)

\bibitem{Mora:2010b}
Mora, T., Walczak, A.M., Bialek, W., Callan, C.G.: {Maximum entropy models for
  antibody diversity.}
\newblock Proc Natl Acad Sci U S A \textbf{107}(12), 5405--5410 (2010)

\bibitem{Nelson:2015kc}
Nelson, R.W., Beisang, D., Tubo, N.J., Dileepan, T., Wiesner, D.L., Nielsen,
  K., W{\"u}thrich, M., Klein, B.S., Kotov, D.I., Spanier, J.A., Fife, B.T.,
  Moon, J.J., Jenkins, M.K.: {T Cell Receptor Cross-Reactivity between Similar
  Foreign and Self Peptides Influences Naive Cell Population Size and
  Autoimmunity}.
\newblock Immunity \textbf{42}(1), 95--107 (2015)

\bibitem{Ninio:1975}
Ninio, J.: {Kinetic Amplification of Enzyme Discrimination}.
\newblock Biochimie \textbf{57}(5), 587--595 (1975)

\bibitem{Qi:2001}
Qi, S.Y., Groves, J.T., Chakraborty, A.K.: {Synaptic pattern formation during
  immune recognition}.
\newblock Proceedings of the National Academy of Sciences \textbf{98}(12),
  6548--6553 (2001)

\bibitem{Rotem:2010}
Rotem, E., Loinger, A., Ronin, I., Levin-Reisman, I., Gabay, C., Shoresh, N.,
  Biham, O., Balaban, N.Q.: {Regulation of phenotypic variability by a
  threshold-based mechanism underlies bacterial persistence.}
\newblock Proceedings of the National Academy of Sciences of the United States
  of America \textbf{107}(28), 12,541--12,546 (2010)

\bibitem{SHANNON:1948}
Shannon, C.E.: {A Mathematical Theory of Communication}.
\newblock Bell System Technical Journal \textbf{27}(3), 379--423 (1948)

\bibitem{Siggia:2013}
Siggia, E.D., Vergassola, M.: {Decisions on the fly in cellular sensory
  systems.}
\newblock Proc Natl Acad Sci U S A \textbf{110}(39), E3704--12 (2013)

\bibitem{Singh:2015vs}
Singh, V., Nemenman, I.: {Accurate sensing of multiple ligands with a single
  receptor}.
\newblock arXiv.org  (2015)

\bibitem{Stefanova:2003}
Stefanov{\'a}, I., Hemmer, B., Vergelli, M., Martin, R., Biddison, W.E.,
  Germain, R.N.: {TCR ligand discrimination is enforced by competing ERK
  positive and SHP-1 negative feedback pathways}.
\newblock Nature immunology \textbf{4}(3), 248--254 (2003)

\bibitem{Taylor:2015}
Taylor, M., Jee, N., Gartner, Z., Mayor, S., Vale, R.D.: {Stimulating T cel
  activation with DNA-based receptors and ligands}.
\newblock In: K.~Symposia (ed.) T cells, regulation and effector function
  (2015)

\bibitem{Tkach:2014}
Tkach, K.E., Barik, D., Voisinne, G., Malandro, N., Hathorn, M.M., Cotari,
  J.W., Vogel, R., Merghoub, T., Wolchok, J., Krichevsky, O., Altan-Bonnet, G.:
  {T cells translate individual, quantal activation into collective, analog
  cytokine responses via time-integrated feedbacks.}
\newblock eLife \textbf{3}, e01,944 (2014)

\bibitem{Tkacik:2011}
Tkacik, G., Walczak, A.M.: {Information transmission in genetic regulatory
  networks: a review.}
\newblock Journal of Physics, Condensed Matter \textbf{23}(15),
  153,102--153,102 (2011)

\bibitem{Torigoe:1998vj}
Torigoe, C., Inman, J.K., Metzger, H.: {An unusual mechanism for ligand
  antagonism}.
\newblock Science  (1998)

\bibitem{Tsitron:2011hq}
Tsitron, J., Ault, A.D., Broach, J.R., Morozov, A.V.: {Decoding complex
  chemical mixtures with a physical model of a sensor array.}
\newblock PLoS Comput Biol \textbf{7}(10), e1002,224--e1002,224 (2011)

\bibitem{Vergassola:2007}
Vergassola, M., Villermaux, E., Shraiman, B.I.: {`Infotaxis' as a strategy for
  searching without gradients}.
\newblock Nature \textbf{445}(7126), 406--409 (2007)

\bibitem{Voisinne:2015}
Voisinne, G., Nixon, G.B., Melbinger, A., Gasteiger, G., Vergassola, M.,
  Altan-Bonnet, G.: {T Cells Integrate Local and Global Cues to Discriminate
  between Structurally Similar Antigens}.
\newblock Cell reports \textbf{11}(5), 1--12 (2015)

\bibitem{Waysbort:2013}
Waysbort, N., Russ, D., Chain, B.M., Friedman, N.: {Coupled IL-2-dependent
  extracellular feedbacks govern two distinct consecutive phases of CD4 T cell
  activation}.
\newblock J Immunol \textbf{191}(12), 5822--5830 (2013)

\bibitem{Youk:2014}
Youk, H., Lim, W.A.: {Sending mixed messages for cell population control}.
\newblock Cell \textbf{158}(5), 973--975 (2014)

\bibitem{Zell:2001}
Zell, T., Khoruts, A., Ingulli, E., Bonnevier, J.L., Mueller, D.L., Jenkins,
  M.K.: {Single-cell analysis of signal transduction in CD4 T cells stimulated
  by antigen in vivo}.
\newblock Proc Natl Acad Sci U S A \textbf{98}(19), 10,805--10,810 (2001)

\end{thebibliography}
\end{document}